\documentstyle[aps,prb,multicol,epsfig]{revtex}

\begin{document}
\input epsf
\draft
\title{Glass transition and layering effects in confined water: \\ 
a computer simulation study}

\author{P.~Gallo$^\dagger$\footnote[1]{Author to whom correspondence 
should be addressed; e-mail: gallop@fis.uniroma3.it
}, M.~Rovere$^\dagger$,  E.~Spohr$^{\ddagger}$}
\address{$\dagger$ Dipartimento di Fisica, 
Universit\`a ``Roma Tre'', \\ Istituto Nazionale per la Fisica della Materia,
 Unit\`a di Ricerca Roma Tre\\
Via della Vasca Navale 84, 00146 Roma, Italy.}
\address{$\ddagger$ Department of Theoretical
Chemistry, University of Ulm, \\ Albert-Einstein-Allee 11, D-89069 Ulm,
Germany }
\maketitle

\begin{abstract}
Single particle dynamics of water confined in a nanopore is studied through
Computer Molecular Dynamics. The pore is modeled to represent the average
properties of a pore of Vycor glass. Dynamics is analyzed at different
hydration levels and upon supercooling. 
At all hydration levels and all temperatures investigated a layering effect
is observed due to the strong hydrophilicity of the substrate. 
The time density correlators show, already at ambient temperature, 
strong deviations from the Debye and the stretched exponential behavior.
Both on decreasing hydration level and upon supercooling we find
features that can be related to the cage effect typical of a 
supercooled liquid undergoing a kinetic glass transition.
Nonetheless the behavior predicted by Mode Coupling Theory can be observed
only by carrying out a proper shell analysis of the density correlators.
Water molecules within the first two layers from the substrate are
in a glassy state already at ambient temperature (bound water). The remaining 
subset of molecules (free water) undergoes a kinetic glass transition;
the relaxation of the density correlators agree with the main
predictions of the theory.  From our data we can predict the
temperature of structural arrest of free water.
\end{abstract}



\begin{multicols}{2}

\section{ Introduction}

The modification of the dynamical properties of liquids
in confined  environment relative to the bulk is 
a field of rapidly growing interest because of the 
close connection with relevant technological and
biophysical problems. 
The specific differences in behavior are, among others, due to
different interactions between liquid and substrate, the size of the
confining region, and the size of the particles composing the liquid.
Nonetheless some underlying common features can be sorted out of the
extremely rich and diversified phenomenology
available.~\cite{grenoble} In fact there are two main causes for the
change of dynamics of a confined liquid with respect to its bulk
phase: 
the first is the influence of the interactions of the liquid with a rough 
surface, which is expected to slow
down the dynamics; 
the second is the confinement effect that can lead to an increase in
the free volume of a molecule, 
which then results in accelerated dynamics together with a decrease of the
glass transition temperature relative to the one of the bulk phase.
The interplay between these two effects depends crucially on 
the particle density
and the size of the confining system.~\cite{Mckenna,kremer1,kremer2,lowen}
 
In particular the dynamical properties of water in restricted
geometries and at interfaces have recently been studied intensely
because of the important effects in systems of interest to biology,
chemistry, and geophysics, whose behavior depends on how the pore
size and structure influence the diffusion of water. Those properties
are particularly relevant in understanding phenomena like the mobility
of water in biological channels~\cite{sansom,lynden-bell} or the
dynamics of hydrated proteins.~\cite{doster}

It is well known that liquid water shows a very peculiar behavior in
the supercooled phase.  The study of water approaching a glass phase
is still a challenging problem since below $235$~K one enters the so
called {\em no man's land }~\cite{stanley}, where nucleation processes
take place and drive the liquid to the solid crystalline phase,
preventing the observation of the glass
transition.~\cite{angell,pablo} It is unclear until now how
confinement could change this scenario. It would be particularly
important to understand whether the glass transition temperature could
be experimentally accessible for confined water.  In this respect the
modifications induced by the confinement on the dynamics of water on
supercooling are of extreme interest.

Computer simulation is a very suitable tool for exploring the liquid
in the range of the supercooled regime without the limitations of the
nucleation process which takes place in the real experiment.
Dynamical properties of different types of liquids in confined
geometries have been studied in recent years by computer
simulation.~\cite{lowen,fehr,boddeker,scheidler} For water confined in
micropores there are a number of computer simulation studies on the
mobility of water.~\cite{sansom,lynden-bell,geiger,gubbins} It is still
difficult, however, to find general trends; systematic studies of the
dynamics of confined water have not been attempted until now.

For confined and interfacial water inelastic neutron scattering and
NMR spectroscopy found a slowing down of the dynamics relative to the
bulk phase.~\cite{chen-nato,chen1,denisov,chen-gallo1,zanotti} 
For water in contact with proteins
there are also signatures of a
transition of adsorbed water to a glassy state, which is 
driven by the protein surface.~\cite{doster} Moreover, 
recent simulation and experimental studies 
showed a typical spectral glassy anomaly, the so called 
boson peak.~\cite{bizzarri1,bizzarri2}

In the many experimental studies of confined water that have been
performed water in Vycor is of particular interest~\cite{chen-nato},
since Vycor is a porous silica glass characterized by a quite sharp
distribution of pore sizes with an average diameter of $40$~\AA. The
pore size does not depend on the hydration level and the surface of
the pore is strongly hydrophilic. Moreover, the water-in-Vycor system
can be considered as a prototype for more complex situations
of interfacial water.~\cite{mar1,mar2}

In this paper we present the results of a Molecular Dynamics study
of the single particle dynamics of water confined in a silica cavity
modeled to represent the average properties of the 
pores of Vycor glass.~\cite{jmliq1,jmliq2} We will concentrate
on the dynamical behavior of the confined water at half hydration on
supercooling. In previous studies the static properties
of this system were investigated.~\cite{jmliq1,jmliq2}   

In the second section we recall the general theoretical background
which concerns the modern interpretation of the glass transition
and we will concentrate on a short description of the 
Mode Coupling Theory (MCT).~\cite{goetze}
In section III we explain the technical details of our simulation
of Molecular Dynamics (MD).
In section IV we discuss the important layering effect that we observe
in the confined water. This effect forms the basis for a layer
analysis of the dynamical results that we introduced for the density
correlator.~\cite{euro} In Section V this analysis is carried out for
the system at ambient conditions.
By means of this layer analysis we can find agreement with
the MCT predictions and estimate the values of some relevant
parameters of the theory as exposed in Section VI where the properties
of the system upon supercooling are analyzed.
Section VII is devoted to the conclusions of our work.

\section{The physics of the glass transition.}

\subsection{The glass transition scenario}

Glasses are a very popular subject of the contemporary science
literature. But in spite 
of the huge experimental, computational and theoretical efforts made, 
a unified picture able to account for the behavior of the liquid from
its normal state throughout supercooling and vitrification is still
lacking.  Theoretical physicists are developing a first principle
theory for glasses~\cite{parisi1,parisi2}, based on the
phenomenological ideas introduced by Kauzmann~\cite{kauzmann} and
developed later by Adam, Gibbs and Di Marzio.~\cite{adam-gibbs}

A scenario emerges that can be pictured by considering
the free energy landscape or alternatively
the potential energy landscape.~\cite{stillinger}
In the high temperature regime the 
free energy functional has only one
relevant minimum and correspondingly 
there are several directions in the configurational space
with negative eigenvalues.
The liquid behaves ``normally'', and 
diffusion is stochastic, i.e., there is only one relevant 
timescale and the relaxation process is of Debye type.  
As one starts supercooling 
(i.e. cooling the system below the melting temperature $T_m$)
the system approaches an important crossover temperature $T_C$ 
referred to as the temperature of kinetic glass 
transition. MCT~\cite{goetze} in its idealized version
is able to describe the dynamics of the liquid in great detail with precise 
predictions on the behavior and the analytical shape 
of the density correlators (see next subsection) for 
systems sufficiently close to, yet above $T_C$.
The region above $T_C$ is only landscape influenced.~\cite{sastry-nature} 
At $T=T_C$ 
the number of directions leading to different basins goes 
to zero.~\cite{cavagna,lanave,donati}
In the region below $T_C$ there is an exponentially large number 
of minima in the free energy separated by barriers that are high 
compared to thermal energy. In this respect $T_C$ represents a 
cross-over temperature from a liquid-like regime to 
a solid-like regime where only hopping processes can restore ergodicity.    
Below $T_C$ the supercooled liquid is in fact frequently trapped in one
of the local free energy minima and only activated processes take 
the system from one minimum to another one in the energy 
landscape.~\cite{adam-gibbs} This region is strongly 
landscape dominated.~\cite{sastry-nature}

In the ``many valley picture'' the liquid-glass transition occurs at
the Kauzmann temperature $T_K$, where
the configurational entropy, which measures the logarithm of the 
number of accessible minima, vanishes and, correspondingly,
the free energy barriers 
diverge.~\cite{parisi1,parisi2,coluzzi,sciort-inerenti} 
The transition at
$T_K$ can be considered as an ideal glass transition which can
take place only at an infinitely slow cooling rate, where it would be
signaled by the divergence of the viscosity.  
This ideal second order transition is
related to the singularities which are found at finite cooling rate in
experiments at the
conventional glass transition temperature, $T_g$, where $T_K < T_g <
T_C$.

\subsection{Predictions of the Mode Coupling Theory}
 
In this paper we will focus on the relatively 
high temperature region where dynamics can be studied by
MD, i.e. the supercooled region where $T$ is above and approaches $T_C$. 
Here MCT in its idealized version works very well 
for many  systems.~\cite{goetze-jcm}

MCT is able to describe the dynamics of a liquid when the single
entity, molecule or atom, is trapped by the transient cage formed by
its nearest neighbors. This transient caging is responsible for the
stretching of the relaxation laws and the separation of time scales.
In its idealized version MCT does not take into account the hopping
processes. This version predicts a transition to a non-ergodic system
at $T_C$, when all the cages are frozen. Nonetheless, hopping
processes are not relevant for most liquids above $T_C$. 
An asymptotic expansion in the region near the ideal singularity (at $T_C$)
yields predictions of the functional form 
of the time correlators and the critical exponents. These predictions
have been tested both by experiments and MD simulations for many glass formers.
MCT thus renders possible, among others, an estimate of the
behavior of the time correlators close to the 
crossover temperature.~\cite{goetze,goetze-jcm}

MCT predicts a two-step relaxation for the dynamics, which
is the signature of the cage effect.  Well above $T_C$, after an
initial ballistic regime, the particle enters the stochastic diffusion
regime, and its mean square displacement (MSD) increases linearly with
time.  In the supercooled region above $T_C$, after the short time
ballistic regime, the particle is trapped by the
barrier created by the nearest-neighbor cage; 
Brownian diffusion is restored only when this cage relaxes.
This behavior is reflected in the 
self part of the density autocorrelation function,
and in any other correlator which has a nonzero overlap with the
density. In particular, when approaching $T_C$ from the liquid side,
the Fourier transform of the density correlator, the self
intermediate scattering function (ISF), $F_S(Q,t)$, has a two step
relaxation behavior with a fast and a slow decay. After the fast decay
it enters a plateau region, corresponding to the rattling of the
particle in the nearest-neighbor cage, which is called the
$\beta$-relaxation region. After the time interval of the plateau,
which becomes longer when approaching $T_C$, the function $F_S(Q,t)$ 
decays to zero. This long time relaxation
is called $\alpha$-relaxation.  In the $\alpha$-relaxation region
it has been shown that the relaxation process is well described by a
stretched exponential
\begin{equation}
F_S \left( Q,t \rightarrow \infty \right) \sim e^{-(t/\tau_l)^{\beta}}
\label{alpha}
\end{equation}
where $\tau_l$ is the long relaxation time and $\beta$ is called 
the Kohlrausch exponent. MCT predicts that, when $T_C$ is approached,
the $\alpha$-relaxation takes place on increasingly longer time
scales, so that the relaxation time $\tau_l$
diverges with a power law
\begin{equation}
\tau_l \sim \left( T - T_C \right)^{-\gamma} \label{taul}\,.
\end{equation}
As a consequence, the diffusion coefficient $D$, which is
predicted to be proportional to $\tau_l^{-1}$, goes to zero
at $T_C$  with the power law $D \sim (T-T_C)^{\gamma}$.

In the idealized version of MCT the system becomes non-ergodic at 
$T_C$, defined as the temperature of structural arrest;
below $T_C$ the cages are frozen and only hopping processes can restore
ergodicity. Extended MCT~\cite{mct-ext} takes into account this effect. 

Finally, we recall that, since the MCT description of the kinetic
glass transition is based on the cage effect, the relevant length
scales are of the order of the nearest-neighbor
distances. Consequently the dynamical quantities in $Q$ space display
this effect most clearly for values of $Q$ close to the maximum of the
static structure factor $S(Q)$.

\subsection{Bulk supercooled water and MCT}

In recent years it was discovered that SPC/E~\cite{spce} supercooled
water has a temperature of structural arrest $T_C$ ~\cite{gallo-prl}
coinciding with the so called singular temperature $T_s$ of Speedy and
Angell.~\cite{SpeedyAngell} As stated above, MCT predicts that close to
$T_C$ the liquid dynamics is dominated
by the ``cage effect''. Water does not behave like normal
glass-forming fluids in this regard, since the cage effect is not a
consequence of an increase of density upon supercooling but rather
seems to be determined by the increase of the hydrogen bond
stiffness, which makes the cage more rigid as the temperature is
lowered below room temperature.  The long time behavior of the single
particle dynamics is well described in terms of the MCT and the
dependence of $\tau_l$ (and $D$) on temperature are found to agree
with the power law, Eq.(\ref{taul}). Successive simulations over a
wide range of pressures and temperatures~\cite{starr} and theoretical
studies~\cite{linda} fully confirmed the MCT behavior of this
potential.

\begin{table}
\caption{Parameters of the SPC/E model and of the substrate-water 
interaction potential.  
\label{tab1}}
 \begin{tabular}{llddd}
    &    & $\sigma$  &$\epsilon/k_B$ &q   \\
    &Site&   $(\AA)$ & (K)       &$|e|$ \\
 \hline
 water\tablenotemark[1]&O & 3.154 & 78.0& -1.04  \\
                       &H & 0.0   & 0.0 & 0.52  \\        
                        &H & 0.0   & 0.0 & 0.52  \\        
 silica\tablenotemark[2]& Si&0.0 & 0.0& 1.283  \\
                        & BO&2.70& 230.0& -0.629 \\ 
                        & NBO&3.00& 230.0& -0.533\\ 
                        & AH &0.0 & 0.0& 0.206 \\
 \end{tabular}
 \tablenotetext[1]{SPC/E model.~\cite{spce}}
 \tablenotetext[2]{Values from Ref.~\onlinecite{brodka}; BO: bridging oxygens;
 NBO: non bridging oxygens; AH: protons on the substrate surface.}
 \end{table}
\begin{table}
  \caption{\label{tab2}
    Hydration levels of the pore. $N_W$ is the number of water
    molecules and $\rho_W$ the corresponding global density.
   The hydration level is based on estimated value for full hydration
   N$_W \sim$ 2600  molecules (see text).}
    \begin{tabular}{ccc}
    N$_W$ & \% hydration & $\rho_W$ ($g/cm^3$) \\ \hline
    500  & 19\% & 0.1687  \\ 
    1000 & 38\% & 0.3373  \\
    1500 & 56\% & 0.4971  \\
    2000 & 75\% & 0.6658  \\
    2600 & 98\% & 0.8788  \\ 
  \end{tabular}
\end{table}
\section{Molecular dynamics of water in Vycor: technical details}

Vycor is a porous silica glass obtained by spinodal decomposition of a
glass-forming melt of SiO$_2$ and B$_2$O$_3$. 
The B$_2$O$_3$-rich
phase is leached out, leaving a SiO$_2$ porous
glass.  
The main differences
of Vycor glass with respect to other analogous porous media are that
it has a well characterized structure with a quite sharp distribution
of pore sizes and a strong hydropylic surface. The void fraction,
$28\%$, is an interconnected network of pores of diameter $\sim
40$~\AA.  
Vycor has a strong capability to absorb water, since its
equilibrium hydration 
at ambient conditions 
is $25\%$ of its dry weight. Furthermore, its pores do not change
size when filled with water.  
For these reasons it can be considered
as a good candidate to study the general behavior of water in
hydrophilic nanopores as a function of the level of hydration.~\cite{mar1}

During experimental sample preparation, after the dessication process
and before introducing water into the pores, those oxygen atoms on the
internal surfaces whose valences are not saturated (called
non-bridging oxygens) are saturated with hydrogen atoms.

In our simulation we build up a cubic cell of silica
glass by the usual procedure of melting a $\beta$-cristobalite 
crystalline structure at $6000$~$K$ and quenching to room temperature.  
As described in details in previous work~\cite{jmliq1,jmliq2} we get 
a cube of length $L=71.29$~\AA. Inside the cube we carve a cylindrical
cavity of $40$~\AA{} diameter by eliminating all the atoms lying within
a distance $R = 20$~\AA{} from the axis of the cylinder taken as the 
{\it z}-axis. Then we exclude all the silicons with less than four
oxygen neighbors. On the surface of the pore 
we distinguish the oxygen atoms which are
bonded to two silicons
(so-called bridging oxygens (BO)) from those which
are bonded to only one silicon atom called non-bridging oxygens (NBO).
The NBOs are saturated with hydrogen atoms.
The final block of material contains $6155$ 
silicon atoms, $12478$ BOs, $227$ NBOs saturated by $227$ acidic
hydrogens (AHs).  The surface density of AH results to be
$2.5$~nm$^{-2}$, in good agreement with the experimentally determined
value of $2.3$~nm$^{-2}$.~\cite{hirama}

Water molecules described by the Simple Point Charge/Extended (SPC/E)
model~\cite{spce} are introduced in the cavity.  The SPC/E potential
is one of the most frequently used interaction site models for water.
Both density and diffusion coefficient show a remarkably good
agreement with experiment at ambient condition.~\cite{spce} It has to
be stressed that SPC/E exhibits a minimum in the pressure P versus T
at about 240 K, while the experimental temperature of maximum density
is located at $T\simeq 277$~$K$.  We therefore expect the phase
diagram of SPC/E water to be shifted downwards in temperature relative
to the experimental one.~\cite{poole,baets}

The water sites interact with the atoms of the rigid matrix by means of
an empirical potential model~\cite{jmliq1,brodka}, where different fractional
charges are assigned to the sites of the silica glass.
In addition, the oxygen sites of water interact with BOs and NBOs of the substrate
via Lennard-Jones potentials. All parameters are 
collected in Table~\ref{tab1}.

The molecular dynamics is performed in the microcanonical ensemble
with a timestep of $2.5$~fs.  Each run was equilibrated via a
coupling to a temperature ``reservoir'' by using the Berendsen method
of velocity rescaling.  Data during the initial equilibration period were
discarded, until monitored quantities like the internal
energy showed no trends with time.

Since the simulations of such a large system are rather time
consuming, we used the shifted force method with a cutoff at $9$~\AA{}
for all interactions.  We checked that the use of a larger cutoff or
Ewald summations does not change the trend of the obtained results, as
discussed in previous work.~\cite{jmliq2} Since we need to store long
trajectories for a large number of molecules, we minimized the disk
space by saving them at 
a logarithmic time step.  
With this choice
the size of trajectories for 2000 molecules lasting $1$~ns is
roughly 1.5 GB.
For the lower temperatures several runs of $1$~ns each were performed.

In the following we will present data for five hydration levels of the
pore at ambient temperature, as described in table \ref{tab2}.  For
the case of roughly half hydration, which corresponds to $N_W=1500$
water molecules, we studied five temperatures, namely
$T=298,270,240,220$ and $210$~K.

\begin{figure}
\centering\epsfig{file=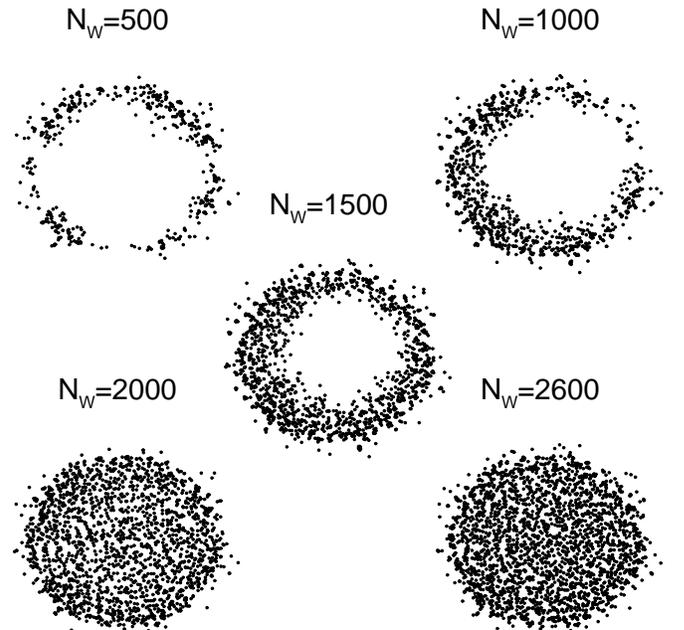,width=1\linewidth}
\caption{Snapshots of equilibrium configurations of confined
water at different hydration levels (see Table II). Only the
oxygen atoms of water are shown projected on the $xy$ plane
perpendicular to the axis of the confining cylinder.
 }
\protect\label{fig:1}
\end{figure} 

\section{Density profiles of confined water}

In the experiments on water confined in Vycor the full hydration is
obtained by immersion of the dry sample in water. Vycor glass adsorbs
water up to $25 \%$ of its dry weight, so the full hydration in the
experiment is defined as $h_f \simeq 0.25$~g~of~water/g~of~Vycor.  A
partial hydration is experimentally obtained by exposing the full
hydrated sample to $P_2 O_5$ for a number of hours in order to reduce
the water content and reach the desired ratio $h$ =
mass~of~water/mass~of~Vycor.~\cite{mar1}

The full hydration in the experiment ($h=h_f$) corresponds to a
density of water~\cite{benham} $\rho_f = 0.8877$~g\,cm$^{-3}$, which
is $11 \%$ lower respect to bulk water.  In computer simulation we
vary the hydration level by changing the number of molecules inserted
in the single cylindrical cavity of our simulation cell. We assume
that the full hydration corresponds to the experimental density
$\rho_f$ 
which would be given in our simulation by a number of water molecules
$N_W \approx 2600$. Levels of hydration are given relative to this
value.

\begin{figure}
\centering\epsfig{file=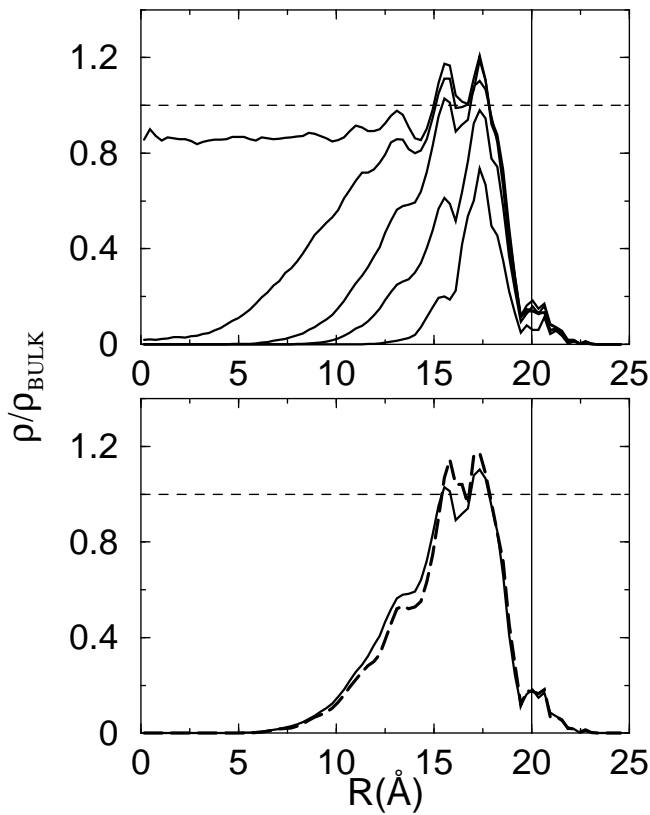,width=1\linewidth}
\caption{Density profiles of confined water along the
cylindrical radius $R=\sqrt{x^2+y^2}$ normalized to the 
bulk water density at ambient temperature. Top:
density profiles at $T=298$ for decreasing levels of hydration
(from $N_W= 2600$ to $N_W=500$ from top to bottom). A small
fraction of water molecules is trapped inside the silica glass,
which leads to the density contribution for $R > 20$ \AA, where the confining
surface is located. 
Bottom: density profiles for $N_W=1500$
at ambient temperature, $T=298$ K, (continuous line) and 
in the supercooled regime, $T=210$ K, (long dashed line).}
\protect\label{fig:2}
\end{figure} 
In Fig.~\ref{fig:1} we show snapshots at ambient temperature for the
five different hydration levels studied, 
$N_W=500$ ($\simeq 20 \%$), $N_W=1000$ ($\simeq 40 \%$),
$N_W=1500$ ($\simeq 60 \%$), $N_W=2000$ ($\simeq 75 \%$) and
$N_W=2600$ ($\simeq 100 \%$), as in Table~\ref{tab2}. 
We observe that the pore is strongly hydrophilic, since molecules
bind strongly to the surface even at low levels of hydration.
The lowest hydration level is lower than 
the monolayer coverage estimated to be $25 \%$ of hydration.~\cite{zanotti}
In the inner surface of the pore single water molecules or 
small patches of water molecules are found.
The radial density profiles normalized to the bulk water density are
shown in Fig.~\ref{fig:2} as a function of $R=\sqrt{x^2+y^2}$.  The
formation of well defined layers of water molecules close to the
substrate is evident in the region between $15$~\AA{} and $20$~\AA{}, 
where the pore surface is located.
For the highest level of hydration water fills the pore completely.
In the pore center a density close to the experimental
density at full hydration is reached.
\begin{figure}[t]
\centering\epsfig{file=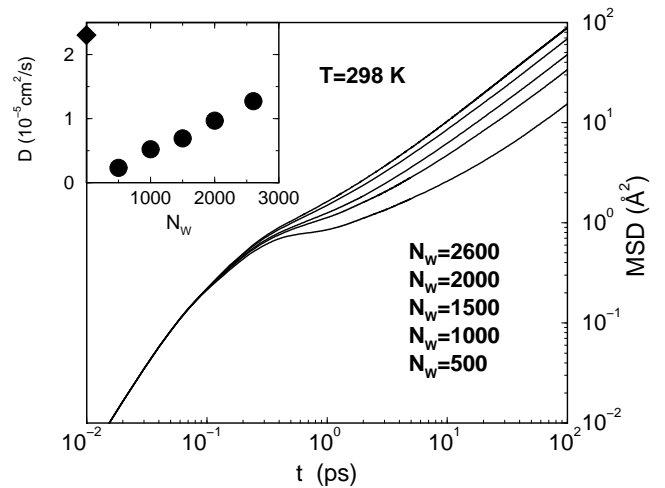,width=1\linewidth}
\caption{Mean square deviation (MSD) of the oxygen atoms 
$<r^2(t)>=< \left| {\mathbf{r}}(t)- {\mathbf{r}}(0) \right|^2 >$
for diffusion in three dimensions at ambient temperature 
as a function of time, $t$,
for different levels of hydration
from top ($N_W=2600$) to bottom ($N_W=500$).
In the inset the diffusion coefficients extracted from
the slopes of the MSD curves 
are displayed as a function of the hydration level (filled circles). 
For comparison the bulk value for SPC/E water at ambient conditions is also shown
(filled diamond).}
\protect\label{fig:3}
\end{figure} 

Layering effects of confined water have been observed in 
almost all computer
simulations performed in different geometry and with different
water-substrate interaction.~\cite{geiger} In our case they are
representative of the high hydrophilicity of the pore. In particular,
we note that, as the hydration level is increased, a double layer
structure with density oscillations is formed. 
Another consequence of the hydrophilic interaction is the distortion
of the hydrogen bond network of water in the layers close to
the substrate.~\cite{jmliq1,jmliq2} Since the inner surface is corrugated,
we do observe that the density is different from zero also right
inside the pore surface, where water molecules are trapped in small
pockets inside the substrate.  These molecules do not contribute to
the diffusion dynamics of water.
A decrease of temperature has no large effect on the density profiles,
as can be seen in the lower part of Fig.~\ref{fig:2} where the case
$N_W=1500$ is shown at ambient temperature ($T=297$ K) and in the
supercooled regime ($T=210$ K). We note that the peak heights of the
two layers become more pronounced as the temperature is lowered.

\section{Slow dynamics of confined water: the role of hydration
at ambient temperature}

We will consider now the single particle translational dynamics of the
oxygens of water as a function of hydration level at ambient temperature.
In Fig.~\ref{fig:3} we display the mean square displacement (MSD)  
$<r^2(t)>=< \left| {\mathbf{r}}(t)- {\mathbf{r}}(0) \right|^2 >$
in three dimensions $<r^2>=<x^2>+<y^2>+<z^2>$
as a function of time for the different hydration levels.
The plot is on a log-log scale
in order to make more evident the flattening of the curves at
lower hydrations. After the initial ballistic regime, where the 
MSD increases proportional to $t^2$ (up to $t \approx 0.4$~ps)
the curves begin to flatten at intermediate times. This behavior
becomes more evident as the hydration level is decreased.
While the onset of the cage effect is almost independent of the hydration,
the relaxation time of the cage depends strongly on the
hydration level. The lowering of the hydration 
affects the mobility of water already at room temperature
by shifting the onset of the diffusive behavior (MSD $\propto t$) to
larger times.~\cite{philmag,euro} 
In the inset the diffusion coefficient, extracted from the slope of
the MSD in the diffusive regime, is plotted as a function of the
number of particles inside the pore. The SPC/E bulk water value is also
displayed. We note substantial decrease of the mobility of confined water 
as the hydration level is lowered.

\begin{figure}
\centering\epsfig{file=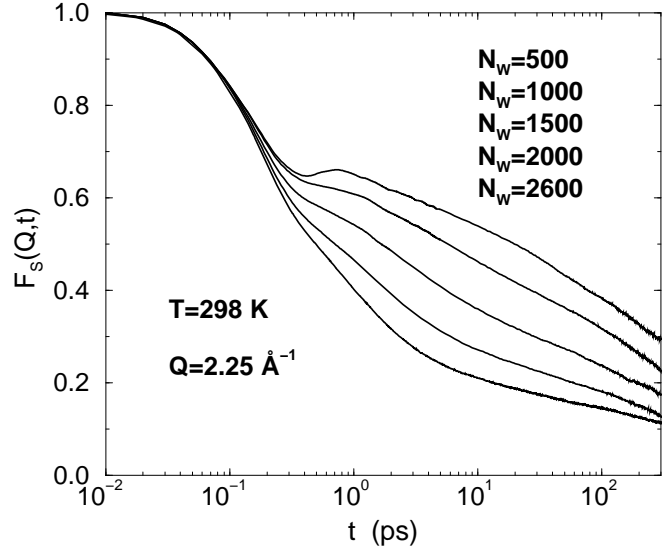,width=1\linewidth}
\caption{Self intermediate scattering function (ISF) of the oxygen
atoms at ambient temperature for $Q=2.25$~\AA$^{-1}$, the position of
the maximum of the oxygen structure factor, for different levels of
hydration from $N_W = 500$ (top curve) to $N_W=2600$ (bottom
curve). $Q$ is averaged over all three dimensions.}
\protect\label{fig:4}
\end{figure}

As can be expected, the different regimes in the diffusion are also reflected
in the behavior of the single particle
intermediate scattering function (ISF), $F_S(Q,t)$. 
This is the Fourier transform of the Van Hove self-correlation
function
\begin{equation}
G_S (r,t) = \frac{1}{N} \left< \sum_{i=1}^{N} \delta \left[
{\mathbf{r}} + {\mathbf{r}}_i(0) - {\mathbf{r}}_i(t) \right] 
\right> \label{grs}
\end{equation} 
$G_S (r,t) d{\mathbf{r}}$ is proportional to 
the probability of finding
a particle at distance $\mathbf{r}$ after a time $t$ if the same particle
was in the origin $\mathbf{r}=0$ at the initial time $t=0$.
The incoherent, or self, ISF can be written as:
\begin{equation}
F_S(Q,t)= \left< \sum_{i=1}^{N}
e^{i{\mathbf Q}\cdot [
{\mathbf r}_i(t)- {\mathbf r}_i(0)]} \right>
\end{equation}
In Fig.~\ref{fig:4} we show the function $F_S(Q,t)$ for the different
hydrations calculated at the maximum of the oxygen-oxygen structure
factor $Q_{max}=2.25$~\AA$^{-1}$, where the cage effect is expected to
be strongest (see discussion in Sec.~II~B).  A shoulder appears in the
correlators upon decreasing the hydration, and the long time
relaxation behavior deviates strongly from the exponential form.  It
is evident that the interaction with the hydrophilic substrate plays a
crucial role in the cage effect, since, at the lower levels of
hydration, the fraction of molecules in contact with the surface
is larger than at higher levels of hydration.

Apparently, lowering the hydration plays a role similar to 
supercooling the bulk liquid, but,    
in spite of some similarity with bulk supercooled water, 
here the late part of the ISFs cannot be fitted by the stretched exponential of Eq.~(\ref{alpha}),
which is predicted by MCT. 

Thus, at first glance our study of confined water appears to indicate
that the water mobility is reduced relative to the bulk phase and
that, as a function of hydration, a diversification of relaxation
times develops analogous to supercooled systems undergoing a kinetic
glass transition.  Nonetheless the deviation for the late part of the
correlator from the analytical predictions of MCT has to be explained
for a full understanding of the dynamics.

We must take into account in our analysis that
the relaxation of the cage around the molecules close to the substrate 
is determined by the strong hydrophilic effect which leads to 
the formation of the double layer structure of water close
to the surface (see Fig.~\ref{fig:2}).  
The way in which the water molecules arrange 
themselves close to the substrate as well as the 
extension of the substrate perturbation affect the
dynamical properties. 

A more detailed examination of the contributions to the single
particle dynamics must therefore be performed by taking into account the
possibility that the
water molecules close to the Vycor surface could be much less mobile
than the ones in the center of the pore, since the former ones  are 
strongly H-bonded to the substrate.~\cite{jmliq1,jmliq2}

For this reason we performed a layer analysis 
separating the contribution coming from the first two layers
(the outer shells or outer layers) close to the substrate
from the contribution coming from the remaining   
water in the inner shells (or inner layers).~\cite{euro} 
The definition of these two subsets is inspired by the 
density profile (see fig.\ref{fig:2}). 
The outer shells are defined as $15<R<20$~\AA{}
while the inner shells are the remaining $0<R<15$~\AA{}.

\begin{figure}
\centering\epsfig{file=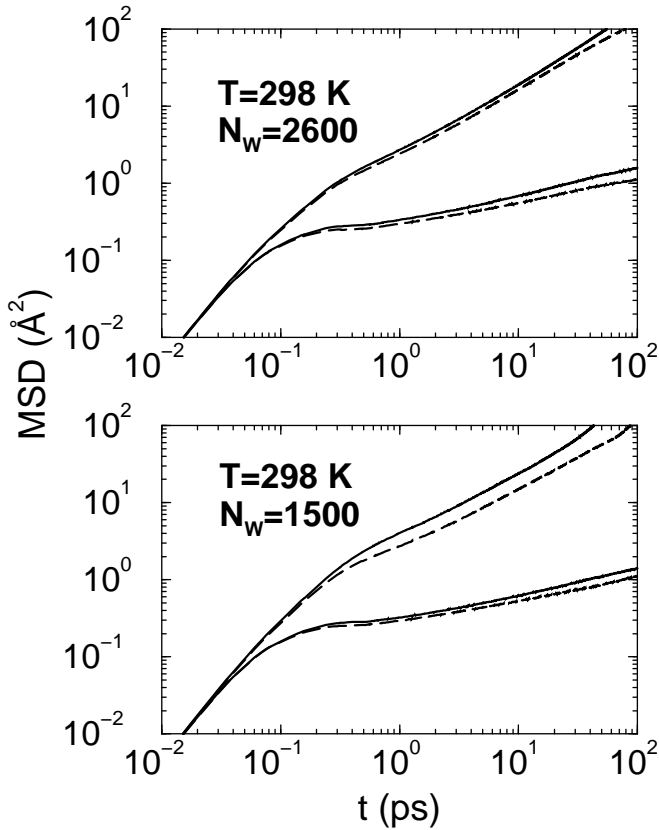,width=1\linewidth}
\caption{Layer analysis of the MSD of the oxygen atoms
at ambient temperature. The 
contributions
of the water molecules in the inner layers (free water) 
and in the outer layers (bound water) are presented 
at ambient temperature for $N_W=2600$ (upper panel) and $N_W=1500$
(lower panel). 
The
continuous lines represent diffusion along the
$z$ direction, while the dashed lines are for the diffusion along the
$xy$ direction.}
\protect\label{fig:5}
\end{figure} 
In Fig.~\ref{fig:5} we show the results of this analysis
for the MSD at two different hydrations, roughly full hydration and
half hydration. 
Since the fluid is confined in the $x$ and $y$ directions we 
calculate separately the quantity along the $z$ direction 
and along the radial (or $xy$) direction  $R$.
Both quantities are multiplied by the proper factor
in order to adjust them to the three dimensional scale
for a direct comparison.

It is clearly  seen that the diffusion of the molecules in the outer 
shells is much slower relative
to the ones in the inner shells. We 
also observe that the change of hydration does not
have a strong influence on the outer shells. 
We find that the diffusion is slower along $xy$ but the shape of
the MSD does not change relative to the one in the $z$ direction.
In fact, the outer water molecules do not reach the diffusive
regime during the observation time of Fig.~5 for both directions.

\begin{figure}
\centering\epsfig{file=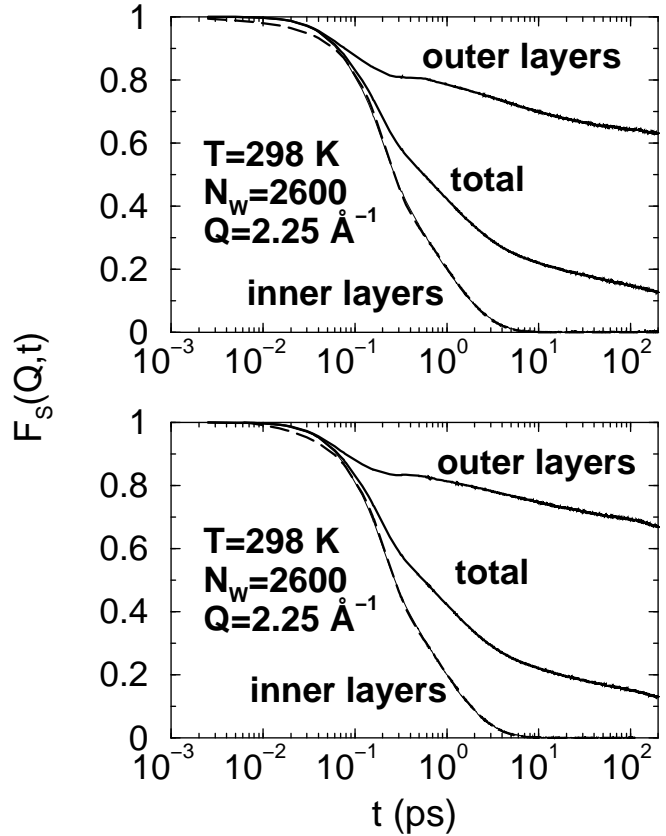,width=1\linewidth}
\caption{Layer analysis of the self intermediate scattering function (ISF) at
ambient temperature for $N_W=2600$ at the peak of the structure factor.
The ISF is shown along the
$z$ direction (top) and along the $xy$ direction (bottom).
In both directions the contributions from free
and bound water are shown separately in addition to the total ISF.
The free water contribution is fitted by Eq.~(\ref{strexp}) (long dashed
line).}
\protect\label{fig:6}
\end{figure} 

The results of the layer analysis for the ISF at full hydration are shown  
in Fig.~\ref{fig:6} along the $z$ direction (top)
and along the $xy$ direction (bottom).
Water in the outer shells appears to be already in a glasslike state,
since its ISF decays to zero over a much longer time scale. The behavior
of the ISF from  the inner layer contribution is different.
We observe that the long time relaxation is characterized 
by a stretched exponential form, Eq.~(\ref{alpha}).
The onset of the $\alpha$ relaxation is preceded by the fast
relaxation, which is  usually observed at short time both experimentally and
in computer simulations of supercooled liquids.
By taking into account the fast relaxation 
with an appropriate normalizing function in both directions
the entire curve can be 
described very well by two relaxation
processes with different time scales 
\begin{equation}
  F_S(Q,t)= \left[ 1-A(Q) \right] e^{-\left( t/\tau_s \right)^2}+
  A(Q)e^{- \left( t/\tau_l \right)^\beta}
\label{strexp}
\end{equation}
where the stretched exponential contains 
the long relaxation time $\tau_l$ and the Kohlrausch exponent $\beta$
already introduced in Sec.~II~B.
In this equation the normalizing function $A(Q)$ is the
Debye-Waller factor, which is also termed the Lamb-M\"ossbauer
factor for the single particle motion.
$A(Q)=e^{-a^2 Q^2/3}$ accounts 
for the cage effect, with $a$ the effective cage radius. The
short time function is written in terms of an exponential
containing the short relaxation time $\tau_s$.  
The values extracted from the fit are reported in table \ref{tab3}.

\begin{figure}
\centering\epsfig{file=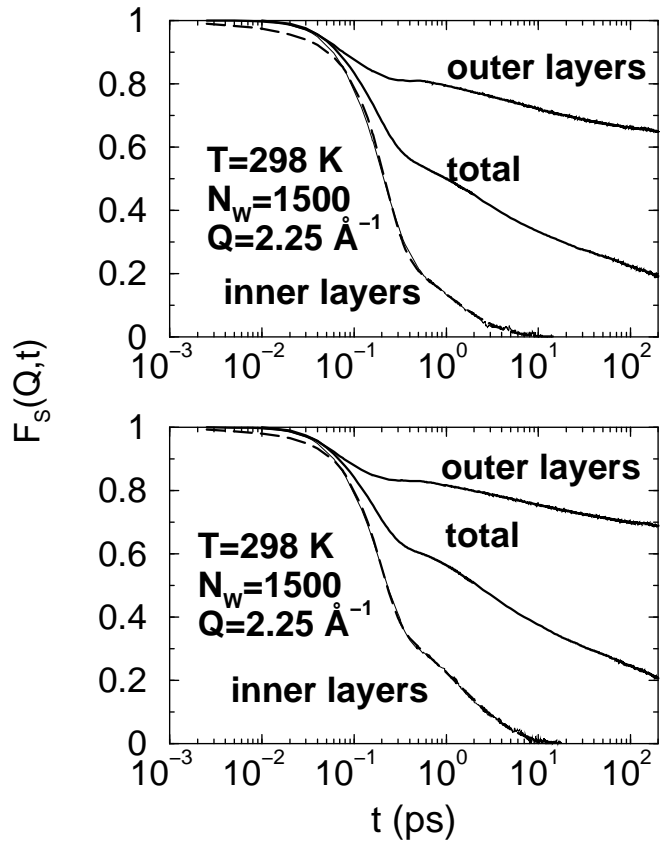,width=1\linewidth}
\caption{Like Fig.~\ref{fig:6} but for $N_W=1500$.\hspace*{\fill} }
\protect\label{fig:7}
\end{figure} 

In Fig.~\ref{fig:7} the shell analysis for the ISF at $Q_{max}=2.25$
\AA{} is shown at half hydration, ($N_W=1500$). The shell analysis
works well also for this level of hydration, and the curves obtained
from the inner layer contribution can be fitted by the functional form
Eq.~(\ref{strexp}).  The resulting parameters are also reported in
table \ref{tab3}.

The $\tau$ parameters reported in table \ref{tab3} are similar to
those of bulk water but $\beta$ is much smaller~\cite{gallo-pccp}
relative to the bulk where $\beta=1$.~\cite{gallo-prl}
From the values of $A(Q)$ the cage radius can be extracted.  We obtain
$a\simeq 0.5$~\AA, again similar to the radius obtained for bulk
water.

We note in passing that if only one layer is considered for the bound
water, namely $17<R<20$~\AA, then the fit to the stretched
exponential of the remaining water is not completely
satisfactory.~\cite{philmag} This confirms that the separation into
the two subsets of water molecules chosen above is an appropriate
representation of the two clearly distinct dynamical regimes coexisting
in the liquid.
It is unclear to what extent these effects would be visible
in the experiments, where usually only the total ISF is measured. There
are, however, experimental indications that two different 
subsets exist in H-bonded confined liquids.~\cite{melni,enza}

We note that water shows an analogous behavior to what was found in mixtures,
for example, with $n$-butoxyethanol. In this case dielectric relaxation
studies show the existence of two kinds of water coexisting in the mixture:
the hydration water  and the ``bulk'' water.~\cite{Fioretto}
Similar behavior has also been reported in 
a theoretical study of water in contact with a
formic acid dimer, where significantly lower values of $D$ have been observed
in the first and second solvation region up 
to $\approx 4.5$~\AA.~\cite{Jorge}
Moreover, the dynamics of a complex liquid, {\it o}-terphenyl (OTP), 
confined in nanometer scale pores 
can also be interpreted in terms of a two layer
model. Calorimetric measurements show in fact the 
existence of two glass transition temperatures for this
latter system.~\cite{Mckenna}

Thermometric 
studies~\cite{takamuku,ewhansen1}, 
NMR spectroscopy~\cite{takamuku,ewhansen1,stapf,ewhansen2}, 
neutron diffraction~\cite{takamuku,dore1,dore2,dore3,mcbf1}, 
and X-ray diffraction~\cite{morishige}
can also be interpreted in terms
of two types of water which are present in confining pores:
{\it free water}, which is in the middle of the pore, and 
{\it bound water}, which
resides close to the surface. Free water is observed to freeze
abruptly in
the cubic ice structure with an hysteresis  
in the melting/freezing transition which decreases at 
decreasing pore size.
Bound water freezes gradually without any hysteris effect
but it does not make any transition to an ice phase and it is
called sometimes {\it nonfreezable} water.~\cite{morishige}
Although all these experimental findings refer to static properties,
we can reasonably identify the water molecules in the
outer layers as {\it bound} and {\it not freezable} water 
while {\it free} water corresponds to the molecules in the inner 
layers of the simulated pore. 
\begin{table}
  \caption{\label{tab3} 
	Fit parameters of the lineshape, Eq.~(\ref{strexp}),
	to the ISF data at ambient temperature.}
    \begin{tabular}{cclll}
    N$_W$ & Direction & $\tau_l$(ps) & $\beta$ & $\tau_s$(ps) \\ \hline
    2600 & xy &  0.83 & 0.78 & 0.22 \\ 
    2600 & z  &  0.84 & 0.79 & 0.21 \\ 
    1500 & xy &  1.4  & 0.67 & 0.19 \\
    1500 & z  &  0.79 & 0.64 & 0.22 \\
  \end{tabular}
\end{table}

\section{Confined water at supercooled temperatures}

\begin{figure}
\centering\epsfig{file=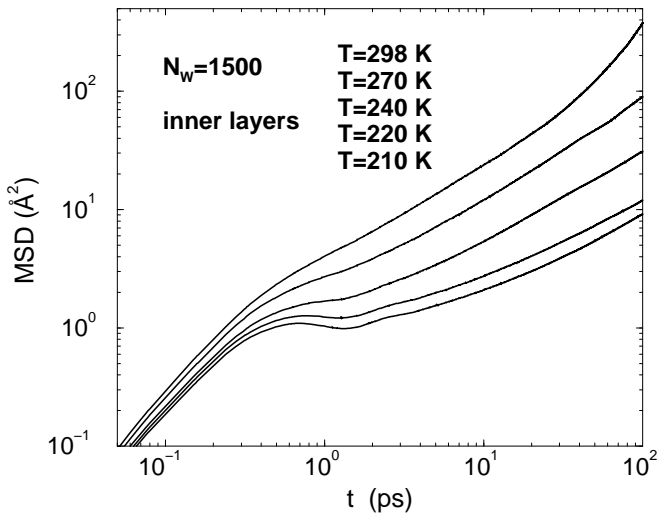,width=1\linewidth}
\caption{MSD of free water at half hydration 
along the $z$ direction for 5 temperatures (298, 270, 240, 220, and 210 K, from top to bottom).
}
\protect\label{fig:8}
\end{figure}

For the discussion of supercooled confined water we consider the case
close to half hydration, $N_W=1500$, which can be regarded as a
reasonable compromise between computational effort and performing the
layer analysis with sufficient statistical accuracy.  As stated above,
the temperature effect on the structure of the double layer of
molecules close to the substrate is small (see
Fig.~\ref{fig:2}). Therefore, the layer analysis is done using the
same separation, $R=15$ \AA{}, as for the room temperature
simulations.~\cite{gallo-pccp} The MSD along the $z$ direction is
shown for decreasing temperatures for the inner shells (free water) in
Fig.~\ref{fig:8} and the outer shells (bound water) in
Fig.~\ref{fig:9}.  The clear distinction between free and bound water
at room temperature is maintained at all lower temperatures. It is
clear from Fig.~\ref{fig:8} that at lower temperatures the cage effect
persists for longer times.

\subsection{The ISF at $Q_{max}$ versus T}

\begin{figure}
\centering\epsfig{file=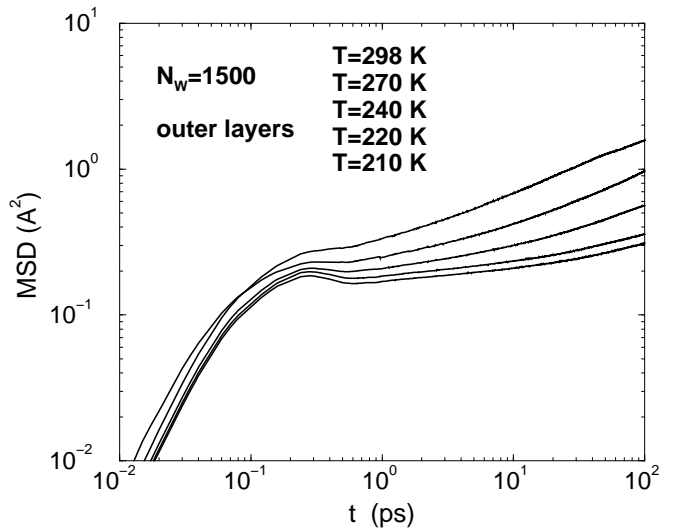,width=1\linewidth}
\caption{MSD of bound water at half hydration 
along the $z$ direction for 5 temperatures (298, 270, 240, 220, and 210 K, from top to bottom).
}\protect\label{fig:9}
\end{figure}

\begin{figure}
\centering\epsfig{file=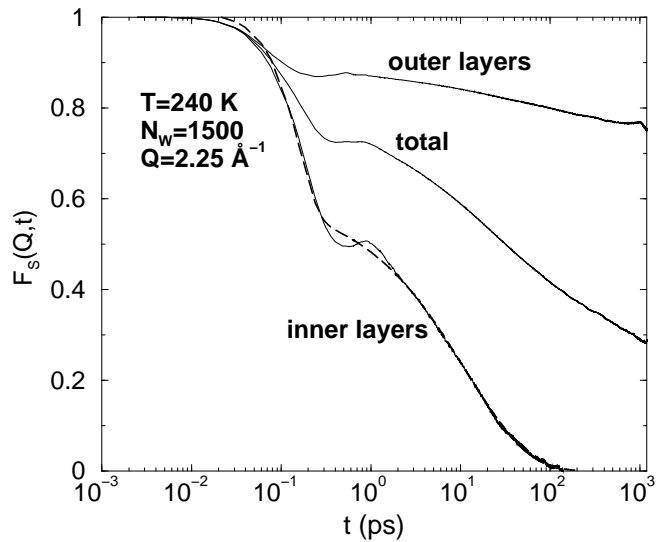,width=1\linewidth}
\caption{Layer analysis of the ISF at $T=240$~K for $N_W=1500$ at
$Q=2.25$~\AA$^{-1}$ along the $z$ direction.  The contributions from
the free (inner layers) and bound water (outer layers) are shown
together with the total ISF.  The free water curve is fitted by
Eq.~(\ref{strexp}) (long dashed line). The bump in the free water ISF
is evidence of the presence of a  Boson Peak.}
\protect\label{fig:10}
\end{figure}

Fig.~\ref{fig:10} shows, as an example, the ISF for motion along the
$z$ direction at $T=240$~$K$ and $Q=Q_{max}=2.25$ \AA{}.  The inner
layer contribution can be fitted very well by Eq.~(\ref{strexp}); we
obtain $\beta=0.62$, $\tau_l=11$~ps, $\tau_s=0.16$~ps.  At this
temperature a bump appears in the ISF of the free water at around
$0.7$~ps. Signatures of this feature are also visible in the MSD of
Fig.~\ref{fig:8} and~\ref{fig:9} where oscillations in the plateau
region are easily discernible. They are similar to features found in
other simulation studies; for typically strong glass formers they
can be attributed to the so called Boson peak
(BP).~\cite{bizzarri1,bizzarri2,horbach1,horbach2}

\begin{figure}
\centering\epsfig{file=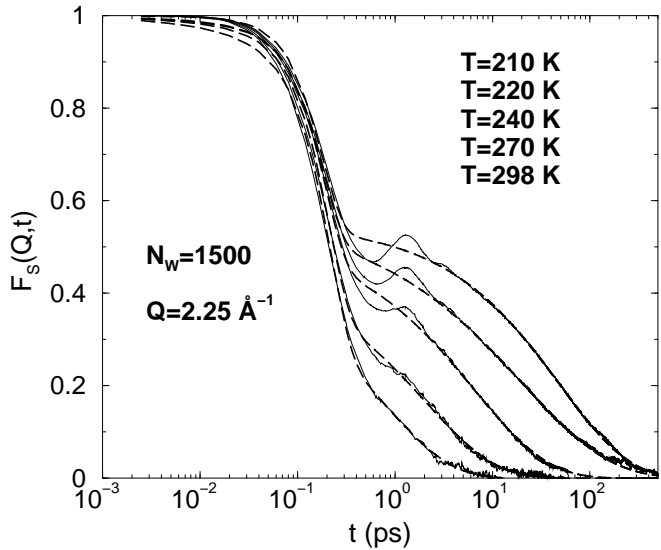,width=1\linewidth}
\caption{The free water contribution to the ISF 
for $Q=2.25$~\AA$^{-1}$ along the $z$ direction
at different temperatures, from $T=298$~K (bottom curve)
to $T=210$~K (top curve).
The curves are fitted by Eq.~(\ref{strexp}) (long dashed lines).
It is evident that the bump, the region of which is excluded
from the fit, increases with decreasing temperature.}
\protect\label{fig:11}
\end{figure}

We repeated the layer analysis of the ISF at $Q_{max}$ for all different 
temperatures studied. 
In Fig.~\ref{fig:11} we show the results along the
$z$ direction for the inner layers as a function of temperature at 
$Q = Q_{max}$. We wish to point out that there are no
qualitative differences in the behavior along the $xy$ direction,
where, in particular, the same BP feature is present.
We observe that, as the temperature is decreased, the shoulder of
the slow relaxation becomes more and more pronounced.
 
The fit by Eq.~(\ref{strexp}) agrees very well in the whole range
explored except for the region of the bump ascribed to the BP,
which is not accounted for in the fitting function.  Since we can
carry out a precise line shape analysis by using only two relaxation
processes, where the long time part is perfectly described by the
$\alpha$ relaxation of the MCT, we can test the predictions of the MCT
which concern the values for the Kohlrausch exponent $\beta$ and the
long relaxation time $\tau_l$.
\begin{figure}
\centering\epsfig{file=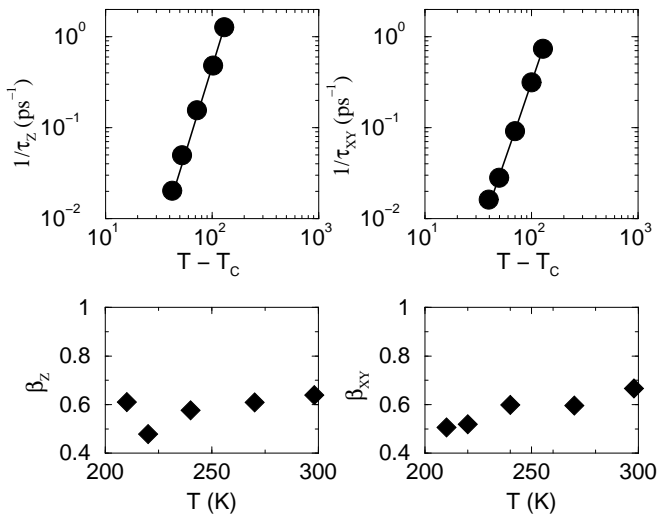,width=1\linewidth}
\caption{Fit parameters.
Top: log-log plots of the inverse relaxation time $\tau_l^{-1}$ as a function of $T-T_C$
along the $z$ direction (left) and the $xy$ direction (right).
The full lines are power law fits given by ${1}/{\tau_z} \sim
(T-167.6)^{3.8}$ (left) and ${1}/{\tau_{xy}} \sim
(T-170.4)^{3.4}$ (right). Bottom:  $T$
dependence of the Kohlrausch exponents $\beta$ along the $z$ (left)
and $xy$ directions (right). }
\protect\label{fig:12}
\end{figure} 
The values of $\tau_l$ and $\beta$
obtained from the fits of the ISF at different temperatures   
are shown in Fig.~\ref{fig:12} for both $xy$ and $z$ directions. 
The validity of the asymptotic scaling law predicted by MCT 
implies that the stretching exponent $\beta$ is temperature
independent.~\cite{goetze-jcm} 
For $T>T_C$ one expects $\beta(T \rightarrow T_C) 
\approx \beta_0 < 1 $. 
In our case the exponent $\beta$ approaches an
asymptotic value around $0.5$ in $xy$ direction
and $0.6$ along $z$ as shown in 
the lower part of Fig.~\ref{fig:12}. 
A similar value is found for bulk water. 
In the panels in the upper part we show that the relaxation time satisfies
the MCT prediction Eq.~(\ref{taul}).
In the $z$ direction  (panel on the left)
we estimate a temperature of structural arrest
$T_C \approx 167.6$~$K$ with an exponent $\gamma \approx 3.8$.
Along the $xy$ direction (panel on the right) 
the $T_C$ is slightly higher $T_C \approx 170.4$~$K$
with $\gamma \approx 3.4$. 

This is in agreement with the idea that an enhancement of the free 
volume speeds up the dynamics leading to a decrease of the glass transition
temperature relative to the bulk. At ambient pressure
SPC/E bulk water undergoes a kinetic glass transition at $T=186.3$~$K$
with $\gamma =2.29$.~\cite{gallo-prl}

\subsection{The ISF at different $Q$}

\begin{figure}
\centering\epsfig{file=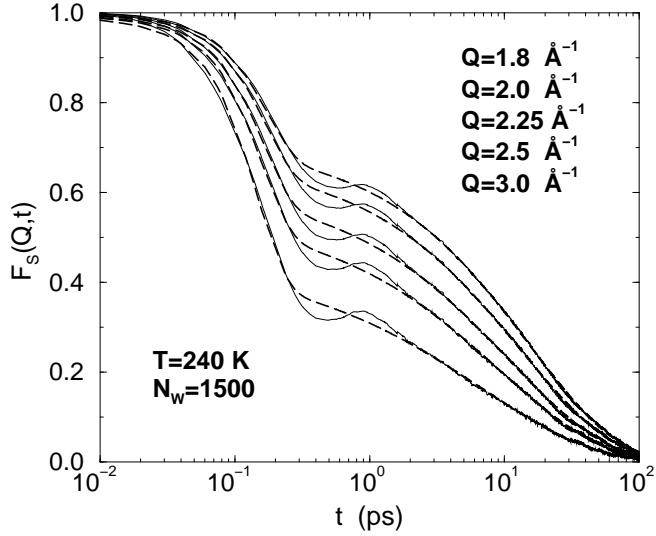,width=1\linewidth}
\caption{ISF along the $z$ direction of the free water 
for half hydration and
$T=240$~$K$ for several values of $Q$ from $Q=1.8$~\AA$^{-1}$ (top curve)
to $Q=3$~\AA$^{-1}$ (bottom curve). All curves are fitted 
by Eq.~(\ref{strexp}) (long dashed lines), excluding the region around the bump.  }
\protect\label{fig:13}
\end{figure}

\begin{figure}
\centering\epsfig{file=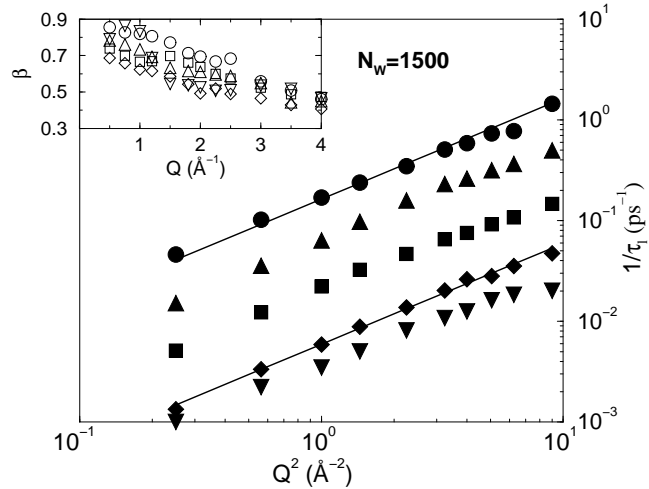,width=1\linewidth}
\caption{Relaxation time $\tau_l$ and exponent $\beta$ (in the inset)
of free water as extracted
from the fits of the long time relaxation behavior of the ISF at 
different $Q$ along the $xy$ direction (Eq.~(\ref{strexp})). 
The continuous lines with slope
$1$ have been inserted as guidelines. 
Different symbols represent different temperatures; temperature
decreases from top to bottom. Open symbols in the inset are
for the same temperature as the corresponding full symbols in the
figure.  }
\protect\label{fig:14}
\end{figure}

\begin{figure}
\centering\epsfig{file=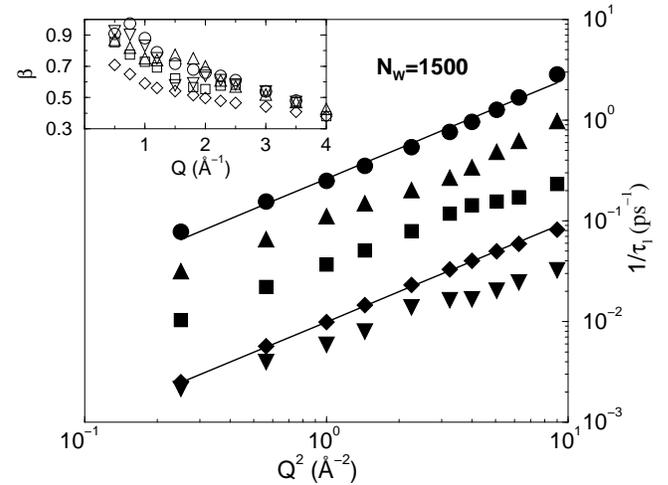,width=1\linewidth}
\caption{Like Fig.~\ref{fig:14}, but along the $z$ direction. }
\protect\label{fig:15}
\end{figure}

From the fits of the ISF at different wavelengths $Q$, shown for example along
the $z$ direction in Fig.~\ref{fig:13} for $T=240$~$K$, we obtain 
values of $\beta$ and $\tau_l$ as a function of $Q$. 
Such values for the different temperatures are reported in Fig.~\ref{fig:14}
for the $z$ direction and in Fig.~\ref{fig:15} for the $xy$ direction.  
The behavior of the parameters is compatible with the MCT.
As shown in the insets of both figures the Kohlrausch exponent
$\beta$ starts from values close to 1 at low $Q$ and high temperatures
and reaches a common plateau at large $Q$.
This behavior is in agreement with the interpretation of the kinetic glass
transition by MCT in terms of
the cage effect, since for $Q^{-1}$ larger than the cage size we 
expect to find that the stretching of the exponential at long
time Eq.~(\ref{alpha}) is less relevant. Thus the Kohlrausch exponent
is expected to go to 1 for large $Q^{-1}$ when the system enters
the normal Debye regime of diffusion
in the long time decay region of the ISF. 
Of course this limit is reached at higher values of $Q$ at higher
temperatures, where the Debye regime is more easily established, as
seen in the insets of Fig.~\ref{fig:14} and~\ref{fig:15}.  The
oscillation in the plateau region does not allow for the calculation
of the MCT parameters a and b related to the power laws of approach to
and departure from the plateau.  It is however shown that for large
$Q$ values the Kohlrausch exponent coincides with the scaling exponent
b.~\cite{Mattias} We therefore obtain $b \approx 0.4$ in this system for both
the $xy$ and the $z$ directions.

$\tau_l^{-1}$ shows a linear behavior
as function of $Q^2$ in Fig.~\ref{fig:14} and~\ref{fig:15},
where the continuous lines which have slope 1 are plotted
as guidelines. This behavior is common for
glass formers close to the kinetic glass transition and has been
found for instance in glycerol.~\cite{glycerol}
In the interpretation we must take into account that 
$\tau_l$ measures the timescale of the
final relaxation of the cage with the subsequent decay of the 
self correlation function of the densities over the length scale $Q^{-1}$   
and the diffusion of the molecule over distances of the order $Q^{-1}$.
In the diffusive regime valid for large $Q^{-1}$ we expect 
$\tau_l^{-1} \sim Q^2$. At variance with bulk water we do not find
any crossover of the behavior of 
$\tau_l^{-1}$ from a $Q^2$ to a $Q$ proportionality 
at high $Q$,~\cite{gallo-prl}
except, perhaps, at the lowest temperature.

In Fig.~\ref{fig:13} there is a clear evidence
of a bump at $0.7$~ps which is preserved at the different temperatures
as seen in Fig.~\ref{fig:11} for $Q_{max}=2.25$~\AA$^{-1}$.
The position of the bump is independent of $Q$, which is clearly seen
in Fig.~\ref{fig:13} at $T=240$~K and which is also found at all 
other explored
temperatures. As mentioned in Sec.~VI~A, 
similar effects have been observed in computer 
simulations of other glass formers~\cite{bizzarri1,bizzarri2,horbach1,horbach2}
and have been attributed to the existence of a Boson peak (BP).    
The BP is 
present in many glasses as a consequence of an excess of vibrational
modes and can be detected even in computer simulation in spite
of the finite size effects.~\cite{horbach1,horbach2} 
We do not want to enter here in a detailed discussion of this behavior,
but we recall that the BP is considered as a precursor of a glass transition
when it appears in the liquid state.
We notice that as seen in Fig.~\ref{fig:10},  the BP feature appears clearly
for the inner shells while it is much less evident in the outer layer 
contribution. 
While in some respect the substrate seems to enhance the effect
relative to bulk water~\cite{gallo-prl}, it appears that water in
the outer shells is not able to sustain the vibrations which are
considered responsible of the overshoot, possibly because we keep
the atoms of the silica substrate rigid during the simulation.  We
intend to discuss this point more extensively in a future paper.

\section{Comparison with experimental data}

The technique that allows for the best comparison with our MD results
is quasielastic neutron scattering (QENS). In fact the self dynamic structure
factor as measured by incoherent QENS is the time Fourier transform of
the ISF. 
The first data appeared in literature on water-in-Vycor were well
fitted with a jump-diffusion
model in a confined medium.~\cite{zanotti-vecchio}
This model was giving reasonable relaxation times, but the size
of the confining region was unrealistic, $5$~\AA.  
An analysis based on MCT of those data was successively 
proposed.~\cite{chen1,chen-gallo1}
The data were also well fitted and the results were consistent with
the appearance upon supercooling of a slow dynamics 
due to the approach to the kinetic glass transition. 
The $Q$ dependence of the inverse of the structural relaxation time
appeared for the $50 \%$ hydration level upon supercooling, 
as a power law with an exponent ranging
from $2.51$ to $2.06$.~\cite{chen-gallo1} 
The fits were however performed on a limited $Q$ range,
approximately from $0.4$ to $1.3$~\AA$^{-1}$.
These findings have been confirmed by a recent 
QENS experiment where a full set of high resolution data 
on water confined in Vycor glass has been analyzed
for different hydration levels upon supercooling.~\cite{zanotti}
The experimental exponents obtained from the fits to a limited
range of $Q$ appear to be slightly different from 
ours. In fact we find, see Figs.~\ref{fig:14}-~\ref{fig:15}, a value which is
approximately $2$ at high temperatures and slightly decreases 
for the lowest temperature. However an analysis performed on
a more extended Q value data, up to $1.8$~\AA$^{-1}$  
showed an exponent which is $2$ at
high temperature and decreases at lower temperatures, similar to
what we found in the present simulation.~\cite{bcdgr}

It is important to notice that the QENS response function
is able to probe only the mobile water, namely the water
above $T_C$, because of experimental resolution limitations.
Therefore the experimental data appear so far compatible with our
findings.


\section{Summary and Conclusions}

In this paper we presented the results obtained by Molecular Dynamics
on the single particle dynamics of SPC/E water confined in a silica
pore, which represents an approximation to the system of water
confined in Vycor. This system can also be considered as
representative of the more general case of water confined in a
hydrophilic nanopore.

Through the study of dynamic time correlators such as the
mean square displacement (MSD) and the self part of the intermediate
scattering factor (ISF), 
we investigated the onset of slow dynamics when 
the hydration level in the pore is decreased or when the liquid is supercooled.

At ambient temperature we found that upon decreasing the hydration
level a flattening of the MSD takes place after the initial ballistic
regime and before the onset of the stochastic diffusion. In
correspondence a two step relaxation behavior appears in the ISF, which
displays a strongly non-exponential decay at long time. In this way
the single particle dynamics has the signatures of the presence of a
cage effect as predicted by Mode Coupling Theory (MCT).

Apparently the effect of hydration is similar to the effect of
supercooling in bulk water, but the long time relaxation of the ISF
cannot be fitted by the stretched exponential predicted by the MCT.
By performing a layer analysis it becomes possible to separate the
contribution to the dynamics of the water molecules close to the
substrate (bound water) from the contribution coming from the
molecules which are in the internal layers (free water). The subset of
molecules belonging to the bound water shows density oscillations and
appears to be in a glassy state with very low mobility already at
ambient conditions. The free water shows a dynamical behavior typical
of glass forming liquids. This behavior can be accounted for by the
idealized MCT in analogy with previous findings on SPC/E water in the
bulk phase.~\cite{gallo-prl,starr,linda}  We would like to stress, 
however, that
in the present work we have considered a much more
complex system than the bulk, where an additional anisotropic effect and a
perturbation by the substrate is present.

The fact that MCT could be used in this framework as a unifying
theoretical approach is highly relevant as a guideline for the systematic
study of the important phenomenology of confined and
interfacial water.

We show evidence that the hydration level together with the
hydrophilicity of the substrate play, through the layering effect, an
important role in determining the dynamical properties
of confined water.  As far as the anisotropy of the system is
concerned, we did not observe large differences in the behavior along
the confined direction ($xy$) and along the pore axis ($z$), where
periodic boundary conditions are applied, apart from the fact that the
mobility along the $xy$ direction is smaller than in the $z$
direction.

By supercooling confined water at half hydration we have found
that the subset of free water undergoes a kinetic glass transition,
which can be described in terms of the MCT.
The layer analysis makes possible the calculation of various
 important quantities related to the glass transition,
like the temperature of structural arrest, $T_C$, and the critical 
exponent $\gamma$. Similar results have been obtained from a
preliminary analysis of the full hydration case.~\cite{prl-nostro} 

MCT has been largely used in the interpretation of the phenomenology
of liquids in the supercooled phase. In the recent attempts to develop
a unified first-principles theory for the glass transition it has been
recognized that MCT keeps its validity in describing the liquid just
above the temperature $T_C$.  It is time to test this theory for
confined fluids, where, in spite of the experimental and computer
simulation efforts, the phenomenology of the dynamics of the confined
and interfacial liquids upon supercooling is still unclear.  We have
shown evidence that the MCT can be applied to the interpretation of
the dynamical behavior of supercooled confined water, if the layering
effects determined by the interaction with the hydrophilic substrate
are carefully taken into account.  We believe that the results
obtained in this work are of great importance in understanding the
phenomenology of the mobility of fluids in nanopores.

\section {Acknowledgements}

P.G. and M.R. are indebted with M.A. Ricci for 
her contribution to the early stages of this work. 
P.G. furthermore thanks Barbara Coluzzi, Daniele Fioretto, 
Mattias Fuchs, Jorge Kohanoff, Friedrich Kremer, Ruth M. Lynden-Bell,
Gregory B. McKenna and Ajay P. Singh for helpful conversations.
We additionally thank Barbara Coluzzi and Francesco Sciortino
for a critical reading of the manuscript.
The financial support of the {\bf G} Section of INFM is 
gratefully acknowledged.

\end{multicols}


\end{document}